Structure and properties of chemically prepared nanographene islands characterized by scanning tunneling microscopy

Mayu Yamamoto, Seiji Obata and Koichiro Saiki*

Department of Complexity Science and Engineering, The University of Tokyo, Kashiwanoha 5-1-5, Kashiwa, Chiba 277-8561, Japan

Abstract

Single layer graphene islands with a typical diameter of several nanometers were grown on a Pt (111) substrate. Scanning tunneling microscopy (STM) analysis showed most of islands are hexagonally shaped and the zigzag-type edge predominates over the armchair-type edge. The apparent height at the atoms on the zigzag edge is enhanced with respect to the inside atoms for a small sample bias voltage, while such an enhancement was not observed at the atoms on the armchair edge. This result provides an experimental evidence of spatially (at the zigzag edge) and energetically (at the Fermi level) localized edge state in the nanographene islands, which were prepared chemically on Pt (111).

*Correspondence: K. Saiki, Department of Complexity Science and Engineering, The University of Tokyo, Kashiwanoha 5-1-5, Kashiwa, Chiba 277-8561, Japan
E-mail: saiki@k.u-tokyo.ac.jp

**Introduction**

Graphene is a single layer of $sp^2$-bonded carbon atoms that are densely packed in a honeycomb crystal lattice. A recent explosive increase in investigation of graphene was ignited by Geim's group in 2004 through their success in isolating a graphene sheet on a $SiO_2$ substrate.[1] Owing to its characteristic band dispersion, charge carriers in graphene behave as massless Dirac Fermions, which has provided various intriguing properties in the field of fundamental physics. Furthermore the very high carrier mobility qualifies graphene as a promising material for future electronic devices. Independently of the current interests, graphene edge has been paid attention to for the last decade by several theoretical physicists. Fujita et al. calculated the electronic states of graphene nanoribbons by tight binding method and found that peculiar localized states originating from the $\pi$ orbitals appear only at the zigzag edge.[2] When the nanoribbon is terminated with zigzag edges, a dispersionless flat band appears at the Fermi level in the Brillouin zone. When the nanoribbon is terminated with armchair edges, on the other hand, such a flat band doesn't appear at all. Although this calculation was done from the energy band picture, molecular orbitals of large polyaromatic hydrocarbons had been known to show a similar result. Stein and Brown calculated $\pi$ orbital states of those molecules and found appearance of finite density of states only at the anthracene-type edge (zigzag edge).[3] Both band and bond pictures, thus, indicate that non-bonding $\pi$ states produce electronic states localized spatially at the edge atoms and energetically at the Fermi level, which are called "edge states".

One of the most striking features of the edge states is possibility of magnetic order, originating from the electron-electron interaction. Fujita et al. estimated that even for small *U*; on-site Coulomb repulsion, the magnetization appears at the edge site, while it doesn't appear in bulk graphite until *U* becomes considerably larger. It should be noted, however, that the ground state is a ferrimagnetic state in which the spin direction is just reverse to each other at the opposite sides of nanoribbon. Kusakabe and Maruyama pointed out that magnetic moment remains for the graphene nanoribbon with asymmetrized edges based on the local-spin-density calculation.[4]

Although very intriguing properties were predicted from the theories, there have been only a limited number of experimental works with respect to the graphene edges and edge states. The edge structure and local electronic state of graphite were investigated by use of scanning tunneling microscope (STM) so far. Kobayashi *et al*. observed a step edge of bulk graphite and found that the zigzag edges were much shorter in length than the armchair edges and were less frequently observed.[5] In their specimens the edge didn't necessarily consist of a single-layer graphene because the observed edge was a step edge of bulk graphite, which appeared only by chance. Niimi *et al.* observed the edge of bulk graphite by exfoliating the bulk graphite.[6] They measured the scanning tunneling spectrum (STS) at the edge and found that an additive peak appeared in the STS only at the zigzag edge. They mentioned, however, that the tunnel current was unstable right on the edge and spatial resolution was insufficient. Thus, the edge structure and its local electronic structure has not been investigated yet without the influence of underlying graphene layers as far as the step edge was prepared from bulk graphite.

Alternative way to observe the graphene edge is fabricating graphene on a metal substrate via chemical vapor deposition. Growth of a single graphene sheet has been examined on several kinds of metal single crystals such as Pt,[7,8] Ni,[9] and Ir.[10] $C_2H_4$, $C_2H_6$ and CO molecules were decomposed and polymerized on the metal substrates at elevated temperatures. In a previous work we found that nanographene islands were formed by exposing a Pt (111) substrate to benzene ($C_6H_6$) gas and annealing subsequently in ultrahigh vacuum.[11] The size of nanographene islands was found to be adjusted by the substrate temperature at which $C_6H_6$ was adsorbed. Ultraviolet photoelectron spectroscopy (UPS) and near-edge x-ray absorption fine structure (NEXAFS) measurements revealed that an additional density of states (DOS) appeared around the Fermi level in both occupied and unoccupied electronic states for the nanographene islands with a diameter of islands less than 5 nm. However, the relation between the additional DOS and zigzag edge was not certain because the atomic structure of the nanographene island was not observed in that work.

In the present work, we characterized the nanographene islands grown on Pt (111) by means of STM, mainly focusing on the edge structure. Based on the atomically resolved STM image, we discuss whether the zigzag edge really appears on the nanographene islands or not. We further explored the localized electronic states at the zigzag and armchair edges by measuring the sample bias voltage dependence of STM image.

**Experimental**

Experiments were carried out in ultrahigh vacuum (UHV) chamber with a base pressure of $10^{-8}$ Pa. Graphene was prepared on Pt (111) by decomposing $C_6H_6$ under UHV. The preparation procedure was described in ref. 11. Briefly, a Pt (111) substrate was cleaned by repeated cycles of $Ar^+$ sputtering and annealing until the atom image of Pt was clearly resolved. The Pt surface was exposed to $C_6H_6$ with a typical dose of 200 L at room temperature (RT) and was annealed at 870 K for 30 minutes after evacuating $C_6H_6$.

STM measurement was done after the Pt substrate was cooled down to RT.

**Results and Discussion**

Figure 1 displays a typical STM image of Pt (111) after the cleaning procedure. Wide terraces of a few tens of nanometers appeared and the atom image corresponding to a Pt (111) surface is observed as shown in the inset. After the CVD process, graphene nanoislands are formed as shown in Figure 2. Most of the islands have a hexagonal shape and seem to grow from the step edge of Pt (111). This was consistent partially with the results of Land et al.,[7] and Fujita et al.[8] In the former report, most of islands grew from the step edge, while the island shape is not necessarily a hexagon.[7] In the latter report, the island shape is mostly a hexagon, while the graphene grew over the step edges.[8] In the present work, the experimental condition was similar to the former report except the agent gas; ethylene ($C_2H_4$)[7] or benzene ($C_6H_6$). In the latter report $C_6H_6$ was decomposed at high temperatures, at which migration of Pt might occur.[8] Thus, a slight difference in shape and growth mode can be ascribed to the difference in experimental conditions.

Figure 3 shows a magnified image of nanographene islands. A honeycomb lattice is observed clearly in the STM image, in which all carbon atoms can be recognized at the sites indicated by circles in the lower part of the figure. Appearance of all carbon atoms in a graphene layer is in contrast to the STM image of bulk graphite, in which only half of carbon atoms are seen like a triangular lattice. This comes from the inequivalence between A and B carbon sites in the bulk graphite, reflecting its stacking nature. Unlike the bulk graphite, the graphene layer on Pt (111) is a single layer so that all carbon atoms appear with the same intensity.

A closer observation of nanographene islands reveals appearance of moiré structure in some islands. Figure 4 shows typical moiré structures superposed on the atomically resolved graphene lattice. There are several kinds of superstructures as listed in Table 1. All the moiré structures could be reproduced by superposing the graphene lattice onto the Pt (111) lattice, although in most structures the graphene lattice needs to be rotated by various angles from the Pt lattice. The fact that the moiré structure can be explained by superposition of two lattices also supports the formation of a single layer graphene on Pt (111). The variety of moiré patterns in the system of graphene on Pt (111) is a great contrast to that of graphene on transition-metals; Ni (111),[9] Ir (111),[10] and Ru (0001)[12,13], in which the orientation of graphene is uniquely determined so as to form a commensurate lattice. In the case of graphene on Pt (111), however, eight different patterns are observed, suggesting a rather weak interaction between the graphene and the Pt surface.

Some of the atomically resolved STM images provide clear atom images of the edge structure. Two typical examples are shown in Figure 5. Fig.5(a) displays a nanographene island surrounded by a zigzag edge. The height profile (Fig.5(c)) taken along the line in Fig.5(a) shows periodicity of 0.24 nm, which corresponds to the length between two carbon atoms along the zigzag edge. Fig.5(b), on the other hand, displays a nanographene islands surrounded by the armchair edge. The dips corresponding to the concave of the armchairs are observed with an interval of 0.42 nm in the height profile (Fig.5(d)). Recently, Wassmann et al. discussed the stability of graphene edge taking account of hydrogen termination.[14] Whether the graphene edge is terminated with hydrogen or not could not be known only from the present STM measurement. However, it is likely that most of edge carbon atoms are bound with hydrogen atoms because the source molecule is benzene in the present case.[15] In order to discuss which appears preferentially at the nanographene edge, zigzag or armchair, precise analysis was carried out for thirty-one islands recorded with an atom resolution. Figure 6 shows the statistics from all thirty-one nanographene islands. It is striking that the zigzag edge seems dominant in the nanographene islands formed on Pt (111). In the case of graphene formed on Ru (0001), appearance of zigzag edge was also mentioned, although precise analysis was not presented.[12]

Recently, appearance of zigzag edge has been observed experimentally, when the graphene network was cut by various means. One example was observed during the etching of the graphene by metal nanoparticles, which formed trenches parallel to the zigzag edge directions.[16] Another example was observed during the reconstruction of chemical bonds, which were intentionally broken by electron bombardment on the graphene.[17] During the restoration process, a zigzag edge was preferentially formed. These two cases commonly relate to the breaking process of chemical bond. In the case of graphene formation via the present CVD process, decomposition of hydrocarbon molecules occurred preceding polymerization. It is likely that the zigzag edge can be stabilized during the process, where decomposition or breaking of chemical bond occur for the $sp^2$ system

The bias voltage dependence of STM image was measured to investigate the electronic structure at the graphene edges. Figure 7 shows the STM images of nanographene island surrounded mostly by zigzag edge. In the case of larger sample bias voltage (0.4 V) the island seems homogeneous as shown in Fig.7(a) and (b). For the small sample voltage (0.1 V), however, the edge region becomes brightest. At the armchair edges, such a prominent change was never observed. In order to evaluate the change in brightness semiquantitatively, the difference in apparent height ($h_1 - h_2$) between the edge atom ($h_1$) and the inside atom ($h_2$) was plotted as a function of the sample bias voltage in Figure 8. The height $h_1$ and $h_2$ were taken at the second and tenth atom from the zigzag edge, respectively. Although two data were obtained from two different zigzag edges, the sharp peak was observed commonly around zero bias voltage. Although the result in Fig.8 is shown only for the positive sample bias voltage, a similar tendency was observed for the negative sample bias voltage. Such an enhancement around the zero bias voltage was not observed for the armchair edge. This peak can be attributed to the additional density of states around the Fermi level. Thus we can conclude that the edge state originating from the zigzag edge can be found experimentally in the nanographene islands formed on Pt (111).

**Conclusions**

Nano-scale graphene islands are formed on Pt (111) by thermally decomposing benzene molecules, which were deposited at room temperature. Formation of a single layer graphene was confirmed by observation of honeycomb lattice and moiré structure. Variety of graphene orientation with respect to Pt (111) indicates weak interaction between graphene and Pt, although Pt plays a catalytic role in formation of graphene. Predominance of zigzag edge was found through statistical analysis of nanographene islands. Increase of apparent height around the zero sample bias voltage is observed only at the atoms on the zigzag edge. The existence of spatially (at the zigzag edge) and energetically (at the Fermi level) localized edge state could be confirmed in the nanographene islands prepared chemically on Pt (111).

**Acknowledgements**

Authors would like to thank Dr. Susumu Ikeda for assistance in STM measurement, Dr. Shiro Entani for the discussion on edge states. This work was supported by a Grant-in-Aid for Scientific research from MEXT, Japan.

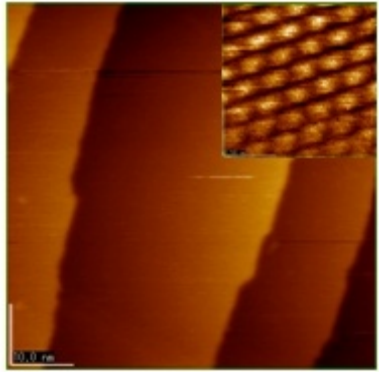

Figure 1. 60 nm x 60 nm STM topograph of a Pt (111) surface. The magnified image (1.5 nm x 1.5 nm) is shown in the inset. The image was taken with a sample bias voltage of $V_s$ =0.1 V and a tunnel current of $I_s$ = 0.4 nA.

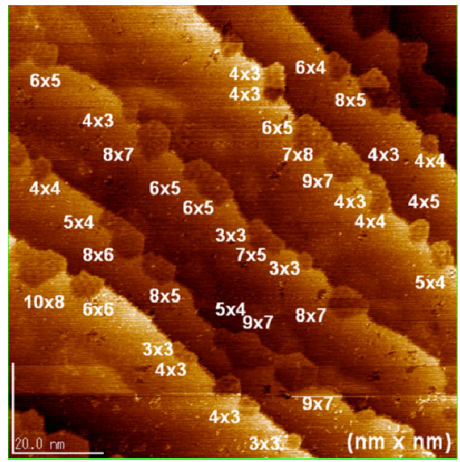

Figure 2. 100 nm x 100 nm STM topograph of nanographene islands grown on a Pt (111) surface. The image was taken with a sample bias voltage of $V_s$ =4.0 V and a tunnel current of $I_s$ = 0.1 nA. Figures denote the size of each island in nm.

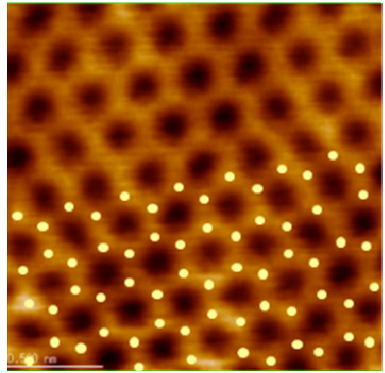

Figure 3. 2 nm x 2 nm STM topograph of nanographene grown on a Pt (111) surface. The image was taken with a sample bias voltage of $V_s$ = 0.004 V and a tunnel current of $I_s$ = 1.0 nA. Circles in the lower part indicate the position of carbon atoms.

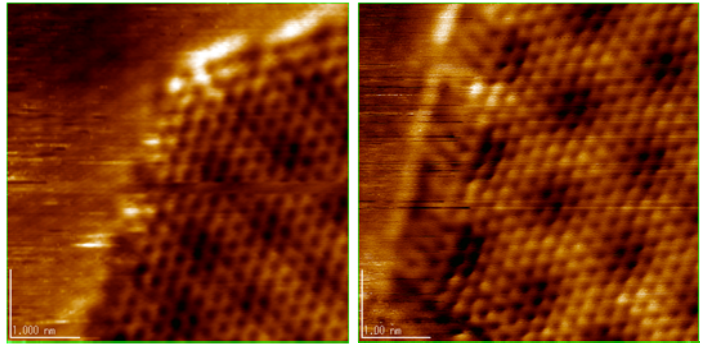

Figure 4. STM topographs of nanographene islands with a superstructure. (left) 4.95 nm x 4.95 nm STM image of graphene/Pt(001) showing a $\sqrt{13} \times \sqrt{13} - R13.9°$ superstructure. The image was taken with a sample bias voltage of $V_s = 0.013$ V and a tunnel current of $I_s = 0.8$ nA. (right) 5 nm x 5 nm STM image of graphene/Pt(001) showing a $\sqrt{21} \times \sqrt{21} - R10.8°$ superstructure. The image was taken with a sample bias voltage of $V_s = 0.008$ V and a tunnel current of $I_s = 1.0$ nA.

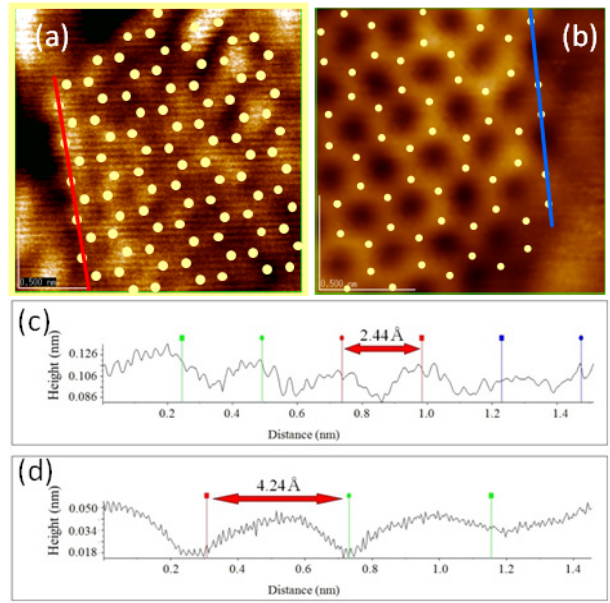

Figure 5. STM topographs of nanographene islands with various edges. (a) Zigzag edge. The image (2 nm x 2 nm) was taken with a sample bias voltage of $V_s$ = 0.497 V and a tunnel current of $I_s$ = 0.975 nA. (b) Armchair edge. The image (1.5 nm x 1.5 nm) was taken with a sample bias voltage of $V_s$ = 0.004 V and a tunnel current of $I_s$ = 1.0 nA. (c) Height profile taken along the line in (a). (d) Height profile taken along the line in (b).

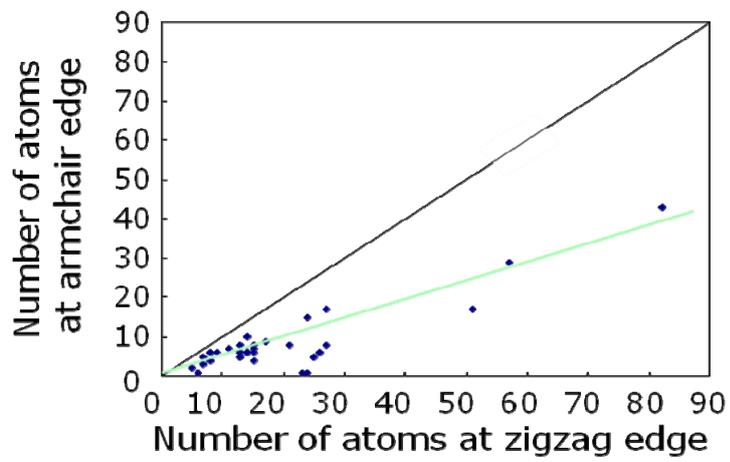

Figure 6. Statistics of thirty-one nanographene islands. Each point corresponds to one island.

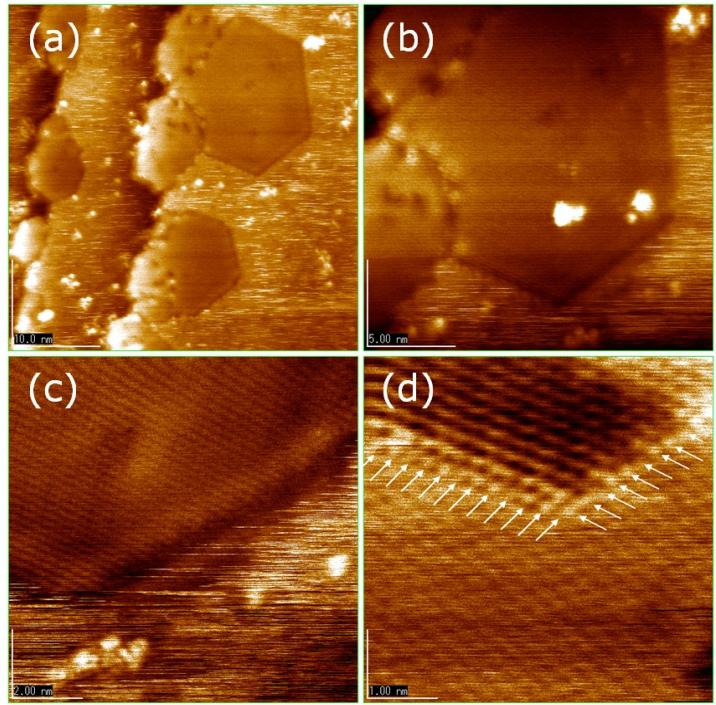

Figure 7. STM topograph of nanographene island at various sample bias voltages ($V_s$). (a) 40 nm x 40 nm, $V_s$ = 0.4 V and $I_s$ = 0.1 nA. (b) 20 nm x 20 nm, $V_s$ = 0.4 V and $I_s$ = 0.1 nA. (c) 10 nm x 10 nm, $V_s$ = 0.2 V and $I_s$ = 0.8 nA. (d) 5 nm x 5 nm, $V_s$ = 0.1 V and $I_s$ = 0.8 nA. <u>The positions of edge atoms are indicated by arrows.</u>

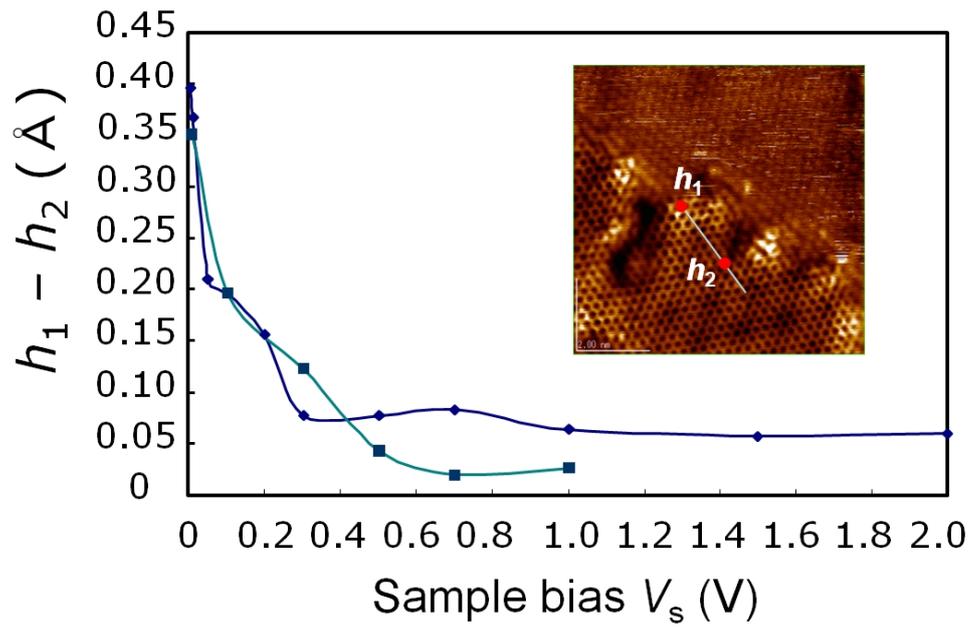

Figure 8. <u>Difference in apparent height</u> between the edge atom ($h_1$) and the inside atom ($h_2$) as a function of sample bias voltage ($V_s$). The STM image was taken at a constant current of 1 nA. The inset shows the position of the edge atom and the inside atom.

Table 1. Superstructures observed in nanographene islands on Pt (111). Periodicity corresponds to the length of unit cell of superstructure.

| Superstructure | periodicity (nm) |
| --- | --- |
| $\sqrt{7}\times\sqrt{7}-R19.1°$ | 0.7 |
| $\sqrt{13}\times\sqrt{13}-R13.9°$ | 1.0 |
| $4\times4-R10.9°$ | 1.1 |
| $\sqrt{19}\times\sqrt{19}-R19.1°$ | 1.2 |
| $\sqrt{21}\times\sqrt{21}-R10.8°$ | 1.3 |
| $\sqrt{37}\times\sqrt{37}-R25.3°$ | 1.7 |
| $\sqrt{48}\times\sqrt{48}-R30°$ | 1.9 |